\documentclass[]{aastex631}
\usepackage{CJK}
\shorttitle{AASTeX v6.31 Sample article}
\shortauthors{Chen et al.}
\graphicspath{{./}{figures/}}
\usepackage{soul}
\usepackage{color,xcolor,tcolorbox}
\usepackage{amsmath}
\usepackage{boondox-cal}
\usepackage[normalem]{ulem}
\usepackage{booktabs,multirow}

\begin{document}
\begin{CJK*}{UTF8}{gbsn}

\title{Theoretical Studies on the Evolution of Solar Filaments in Response to New Emerging Flux}

\author[0000-0002-8077-094X]{Yuhao Chen}
\affiliation{Yunnan Observatories, Chinese Academy of Sciences, P.O. Box 110, Kunming, Yunnan 650216, People's Republic of China}
\affiliation{University of Chinese Academy of Sciences, Beijing 100049, People's Republic of China}
\affiliation{Yunnan Key Laboratory of Solar Physics and Space Science, Kunming, Yunnan 650216, People's Republic of China}
\affiliation{Harvard-Smithsonian Center for Astrophysics, 60 Garden Street, Cambridge, MA 02138, USA}

\author[0000-0001-9828-1549]{Jialiang Hu}
\affiliation{Yunnan Observatories, Chinese Academy of Sciences, P.O. Box 110, Kunming, Yunnan 650216, People's Republic of China}
\affiliation{University of Chinese Academy of Sciences, Beijing 100049, People's Republic of China}
\affiliation{Yunnan Key Laboratory of Solar Physics and Space Science, Kunming, Yunnan 650216, People's Republic of China}

\author[0000-0002-1264-6971]{Guanchong Cheng}
\affiliation{Yunnan Observatories, Chinese Academy of Sciences, P.O. Box 110, Kunming, Yunnan 650216, People's Republic of China}
\affiliation{University of Chinese Academy of Sciences, Beijing 100049, People's Republic of China}
\affiliation{Yunnan Key Laboratory of Solar Physics and Space Science, Kunming, Yunnan 650216, People's Republic of China}

\author[0000-0002-5983-104X]{Jing Ye}
\affiliation{Yunnan Observatories, Chinese Academy of Sciences, P.O. Box 110, Kunming, Yunnan 650216, People's Republic of China}
\affiliation{Yunnan Key Laboratory of Solar Physics and Space Science, Kunming, Yunnan 650216, People's Republic of China}
\affiliation{Center for Astronomical Mega-Science, Chinese Academy of Sciences, Beijing 100012, People's Republic of China}
\affiliation{Yunnan Province China-Malaysia HF-VHF Advanced Radio Astronomy Technology International Joint Laboratory, Kunming, Yunnan 650216, People's Republic of China}

\author[0000-0001-9650-1536]{Zhixing Mei}
\affiliation{Yunnan Observatories, Chinese Academy of Sciences, P.O. Box 110, Kunming, Yunnan 650216, People's Republic of China}
\affiliation{Yunnan Key Laboratory of Solar Physics and Space Science, Kunming, Yunnan 650216, People's Republic of China}
\affiliation{Center for Astronomical Mega-Science, Chinese Academy of Sciences, Beijing 100012, People's Republic of China}

\author[0000-0002-9258-4490]{Chengcai Shen}
\affiliation{Harvard-Smithsonian Center for Astrophysics, 60 Garden Street, Cambridge, MA 02138, USA}

\author[0000-0002-3326-5860]{Jun Lin}
\affiliation{Yunnan Observatories, Chinese Academy of Sciences, P.O. Box 110, Kunming, Yunnan 650216, People's Republic of China}
\affiliation{University of Chinese Academy of Sciences, Beijing 100049, People's Republic of China}
\affiliation{Yunnan Key Laboratory of Solar Physics and Space Science, Kunming, Yunnan 650216, People's Republic of China}
\affiliation{Center for Astronomical Mega-Science, Chinese Academy of Sciences, Beijing 100012, People's Republic of China}

\correspondingauthor{Jun Lin}
\email{jlin@ynao.ac.cn}




\begin{abstract}
New emerging flux (NEF) has long been considered a mechanism for solar eruptions, but detailed process remains an open question. In this work, we explore how NEF drives a coronal magnetic configuration to erupt. This configuration is created by two magnetic sources of strengths $M$ and $S$ embedded in the photosphere, one electric-current-carrying flux rope (FR) floating in the corona, and an electric current induced on the photospheric surface by the FR. The source $M$ is fixed accounting for the initial background field, and $S$ changes playing the role of NEF. We introduce the channel function $C$ to forecast the overall evolutionary behavior of the configuration. Location, polarity, and strength of NEF governs the evolutionary behavior of FR before eruption. In the case of $|S/M|<1$ with reconnection occur between new and old fields, the configuration in equilibrium evolves to the critical state, invoking the catastrophe. In this case, if polarities of the new and old fields are opposite, reconnection occurs as NEF is close to FR; and if polarities are the same, reconnection happens as NEF appears far from FR. With different combinations of the relative polarity and the location, the evolutionary behavior of the system gets complex, and the catastrophe may not occur. If $|S/M|>1$ and the two fields have opposite polarity, the catastrophe always takes place; but if the polarities are the same, catastrophe occurs only as NEF is located far from FR; otherwise, the evolution ends up either with failed eruption or without catastrophe at all.
\end{abstract}



\section{Introduction} \label{sec:intro}

Solar eruptive events, including flares, eruptive filaments, and coronal mass ejections (CMEs), are considered distinct components or different manifestations of the same eruptive process \citep{1995A&A...304..585H,2001ApJ...559..452Z,2004NewA....9..611L,2010hssr.book..159F}. The electromagnetic radiation, high-energy particles, and magnetized plasma generated during the eruptions can severely disturb Earth's magnetosphere and upper atmosphere, leading to disruptions in satellites, damage to terrestrial power systems, and even posing a threat to human health \citep{2021LRSP...18....4T}. Therefore, understanding the mechanisms of solar eruptions is crucial for forecasting hazardous space weather and anticipating its potential adverse impacts on human beings.

Generally, a solar eruptive event occurs as a result of the fast release of the non-potential energy stored in the coronal magnetic field beforehand \citep{2000JGR...10523153F}. The storage of this energy takes place because of the gradual motions of the plasma in the photosphere that continuously deforms the coronal magnetic structure anchored to the photosphere. This stage usually lasts tens of hours, days, or even weeks \citep[][]{2012ApJ...748...77S,2018SSRv..214...46G}. The eruption happens as the coronal magnetic configuration loses equilibrium. This is associated with the consequent reconnection process at a reasonably fast rate; if the rate of reconnection is not high enough, on the other hand, a failed eruption would be expected (see \citealt{2000JGR...105.2375L} and \citealt{2002ChJAA...2..539L} for detailed discussions in theory, and \citealt{2003ApJ...595L.135J} for true events).

The mechanisms driving solar eruptions have been extensively studied (see \citealt{2000JGR...10523153F}, \citealt{2003NewAR..47...53L}, \citealt{2011LRSP....8....1C}, and references therein). Among these mechanisms, the NEF has been recognized as an important triggering mechanism, gaining significant attention over the decades. The NEF is the ultimate cause of the formation of solar active regions \citep{2014LRSP...11....3C}, and occurs at high frequency in both time and space, which plays a crucial role in large-scale eruptions \citep[e.g.,][]{2000ApJ...545..524C,2001JGR...10625053L,2012ApJ...760...31K,2023ApJ...950L...3C}. A recent numerical experiment in the framework of the radiative magnetohydrodynamics suggested that the continuous emergence of magnetic flux may be the origin of the super-hot coronal plasma \citep{2024NatAs.tmp...62L}. Moreover, NEF also drives small-scale activities, such as the chromospheric anemone jet \citep{2007Sci...318.1591S,2018RAA....18...45Z,2021A&A...646A..88N,2022A&A...665A.116N}, the ultraviolet burst \citep{2021RAA....21..229C}, the formation of light bridges in sunspots \citep{2021RAA....21..144L,2024ApJS..271...34L}, and null-point reconnection coupled with mini-filament eruption \citep{2023NatCo..14.2107C}. 

A fundamental question arises: which type of the NEF triggers the large-scale eruption? \cite{1995JGR...100.3355F} explored this issue by analyzing 53 eruptive filament events, and found that when the orientation of the emerging magnetic field are reconnection-favorable with the pre-existing magnetic structure, the eruption is triggered. Through 2.5D MHD simulations, \cite{2000ApJ...545..524C} proposed that magnetic reconnection between NEF and the pre-existing magnetic configuration, which includes a FR, serves as the triggering mechanism for the eruption. Their results support the conclusion of \cite{1995JGR...100.3355F}. Subsequently, \cite{2012NatPh...8..845R} conducted self-consistent 3D numerical simulations, and found that the situation in the true corona could be much more complex than what \cite{2000ApJ...545..524C} displayed. The interaction between NEF and the pre-existing magnetic field results in the formation of a pair of J-type FRs, constituting an S-shape structure that is usually seen in the soft X-ray emission prior to the major eruption.

The conditions under which NEF can drive the coronal magnetic structure to lose the equilibrium can be further understood by applying the principles of catastrophe theory. The catastrophe theory was first introduced by \cite{thom1972stabilite}, and further elaborated by \cite{poston1978catastrophe} to study the stability of structures, such as cells and tissues during their development. Subsequently, \cite{1991ApJ...373..294F}, \cite{1993ApJ...417..368I}, \cite{1994SoPh..150..245F}, and \cite{1995ApJ...446..377F} applied this theory to investigate the evolution of coronal magnetic configurations in response to slow changes in the photosphere. Notably, \cite{1978SoPh...59..115V} also explored the equilibrium and the instability in coronal magnetic configurations even though they did not explicitly use the terms ``loss of equilibrium'' or ``catastrophe.'' Their work constitutes the basis of the catastrophe theory for coronal mass ejections (CMEs). 

On the basis of these works, \cite{2001JGR...10625053L} investigated the quasi-static evolution of the FR as a NEF appears nearby. Their primary purpose is to examine the conclusion drawn from observations by \cite{1995JGR...100.3355F} and that from numerical experiments by \cite{2000ApJ...545..524C}. By presenting several specific cases, \cite{2001JGR...10625053L} pointed out that the situation in reality should be more complex. In fact, whether NEF could trigger the eruption depends not only on its orientation but also on its location, strength, and the way it occurs. Even if the NEF does not reconnect with the existing magnetic structure, it can still lead to the catastrophe. However, \cite{2001JGR...10625053L} did not draw a universal rule or conclusion regarding the role of NEF in triggering the eruption, and it could be considered as a sort of case study.

On the basis of \cite{2001JGR...10625053L}, \cite{2022ApJ...933..148C} further investigated the dynamic evolution after the loss of equilibrium of the FR. Their results showed that the NEF with a reconnection-favorable orientation triggers the catastrophe but may still end up with a failed eruption. Three-dimensional numerical experiments by \cite{2012ApJ...760...31K} highlighted the impact of the shearing angle of the background field and the angle between the NEF and PIL on the final evolution of the FR. While  \cite{2012ApJ...760...31K} demonstrated that shearing motion could also form complex magnetic structures, the NEF remains a common mechanism frequently observed to lead to eruptions. In general, both theoretical studies and observational evidences \citep[e.g., see][]{2008ChA&A..32...56X,2008SoPh..250...75Z} suggested that the criteria governing the eruption triggered by the NEF are highly complex, and currently no simple rule exists.

We note here that, in reality, the NEF is not the only mechanism that may drive the coronal magnetic structure to evolve toward a loss of equilibrium. The case exists that the loss of equilibrium is triggered by the other mechanisms, and the NEF just plays a minor role in this process. On the other hand, the case exists as well that the NEF plays the main role of in triggering the loss of equilibrium. In the present work, we are focusing on the case that NEF is the sole mechanism for triggering the loss of equilibrium.

To establish a general criterion for determining whether NEF can drive an eruption or not, it is essential to further reveal the various effects of different NEFs on the FR configuration. In Section \ref{sec:model}, we introduce the magnetic configurations, including both the FR and the NEF. Section \ref{sec:1} details the initial stage of flux emergence and resulting evolutions of system. In Section \ref{sec:2}, we explore the quasi-static evolution and the catastrophic process as the magnetic flux continuously emerges. Section \ref{sec:3} discusses the dependence of the CME precursor on the NEF. Finally, Section \ref{sec:Conclusion} summarizes this work.

\section{descriptions of the model} \label{sec:model}

\begin{figure}[ht]
\begin{center}
\includegraphics[width=0.4\textwidth]{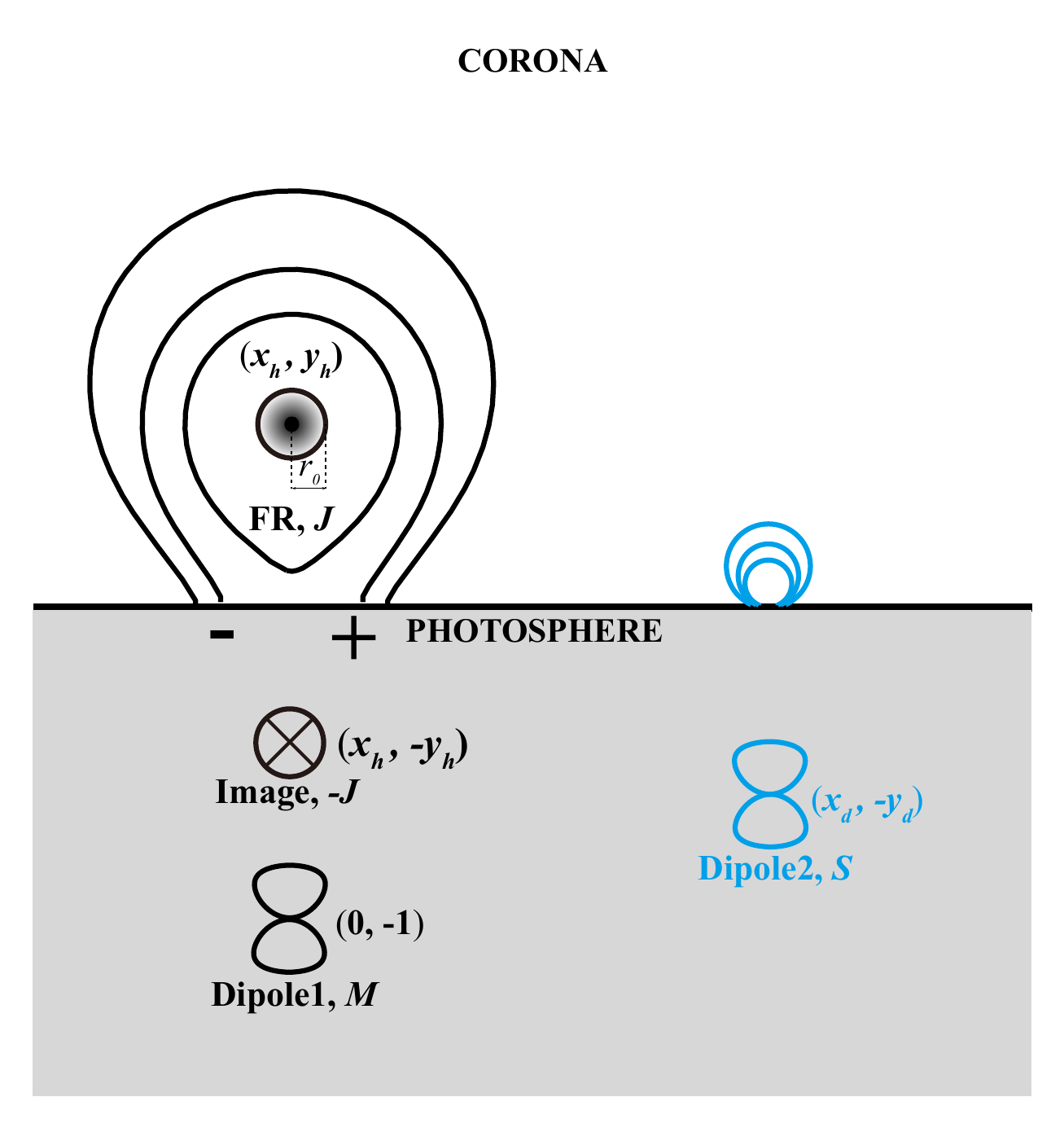}
\end{center}
\caption{The magnetic configuration, including the FR, the image of the FR, the Dipole 1 to model the unchanged background field, and the Dipole 2 to model the NEF. Prior to the magnetic emergence ($S=0$), the FR remained undisturbed, positioning itself just above Dipole 1.
\label{fig:MagCon}}
\end{figure}

Because the magnetic field in the corona dominates the gas pressure and the gravity, and the plasma in the corona possesses fairly high conductivity, we neglect the gravity, gas pressure, and the diffusivity in the present work. The magnetic configuration in this study is described in Figure \ref{fig:MagCon}. The sources of the magnetic field in the corona include Dipoles 1 and 2, the electric current inside the FR, and the surface current induced on the surface of the photosphere, of which the contribution to the coronal magnetic field is equivalent to that of the mirror image of the FR current inside the photosphere. The image current and the current inside the FR have the same intensity, but opposite direction, and have the same distance to the photospheric surface. Dipole 1 located below the photospheric surface contributes to the basic background magnetic field in the corona, which remains unchanged in the evolution of interest; Dipole 2 plays the role in producing the NEF. The interaction between the FR and the image current results in a magnetic compression, pushing the FR away from the Sun (see 
\citealt{1991ApJ...373..294F}; \citealt{1993ApJ...417..368I}; \citealt{1994SoPh..150..245F}).

We set up a Cartesian coordinate system for this work with the $x$-axis on the surface of the photosphere, and the $y$-axis pointing upward. Dipole 1 has strength of $m$ and is located at $(0, -d)$ with both $m$ and $d$ being fixed for the basic background field. Dipole 2 has strength of $s$ and is located at $(x_{d}, y_{d})$, and both its strength and location are changeable. The FR carries the electric current of intensity, $I$, and is located at $(x_{h}, y_{h})$; and its mirror image current of intensity, $-I$, is located at $(x_{h}, -y_{h})$.

Following the practice of \cite{2001JGR...10625053L}, we replace $m$, $s$, and $I$ with dimensionless parameters $M$, $S$, and $J$ in our calculations below via relations $MI_{0}=mc/(4d)$, $SI_{0}=sc/(4d)$, and $I=I_{0}J$. Here $I_{0}$ is a parameter in unit of the electric current intensity, and $c$ is the speed of light. All the lengths in this work are normalized to $d$, and $M$ is set to the unity for simplicity. Therefore, Dipole 1 has strength 1 and is located at $(0, -1)$ in our calculations below. The radius of the FR is denoted as $r_0$.

In the framework set up in this way, $S=0$ corresponds to the case of a single source that has been studied extensively previously (e.g., see \citealt{2003NewAR..47...53L} and \citealt{2006SSRv..123..251F} for a brief review), $S>0$ is the case in which the polarity of the NEF is same as the existing field, and $S<0$ is the case of the opposite polarity. Because of the symmetry in the system of interest, we just need to investigate the case in which $S$ appears at $(x_{d}, -y_{d})$ with both $x_{d}$ and $y_{d}$ positive (see also Figure \ref{fig:MagCon}) in this work.

With the magnetic configuration of interest sketched in Figure \ref{fig:MagCon}, the magnetic field contributed by the sources outside the FR is described in equations (\ref{Bex}) and (\ref{Bey}), and the external force acting on the FR is given in equation (\ref{F}). Following the practice of \cite{2001JGR...10625053L}, setting the external force to zero results in equations (\ref{externalx_eq}) and (\ref{externaly_eq}) for the global equilibrium in the system, the internal equilibrium of the FR leads to equation (\ref{internal_eq}), and the frozen-in flux condition on the surface of the FR gives equation (\ref{forzenflux_eq}). Solving equations (\ref{externalx_eq}) through (\ref{forzenflux_eq}) comprehensively determines the equilibrium locations of the FR in a space spanned by $(S, x_{d}, y_{d}, x_{h}, y_{h})$, and all the equilibrium locations constitute a hyper-equilibrium surface in the space $(S, x_{d}, y_{d}, x_{h}, y_{h})$.

As \cite{2001JGR...10625053L} pointed out that equations (\ref{externalx_eq}) and (\ref{externaly_eq}) could also deduced by taking derivatives of the free energy, $E_{h}$ (see equation \ref{Eh}), in the system with respect to $x_{h}$ and $y_{h}$, respectively. Here the free energy is the difference between the total energy in the system and the energy in the corresponding potential system. By further evaluating
\begin{equation}
\Delta = \left( \frac{\partial^2 E_h}{\partial x_h^2} \right) \left( \frac{\partial^2 E_h}{\partial y_h^2} \right) - \left( \frac{\partial^2 E_h}{\partial x_h \partial y_h} \right)^2, \label{Delta}
\end{equation}
and setting $\Delta =0$, we are able to locate the critical points on the hyper equilibrium surface, at which the catastrophe takes place.

\section{Evolution in the system as the New Flux starts to emerge} \label{sec:1}

\begin{figure}[ht]
\begin{center}
\includegraphics[width=1.0\textwidth]{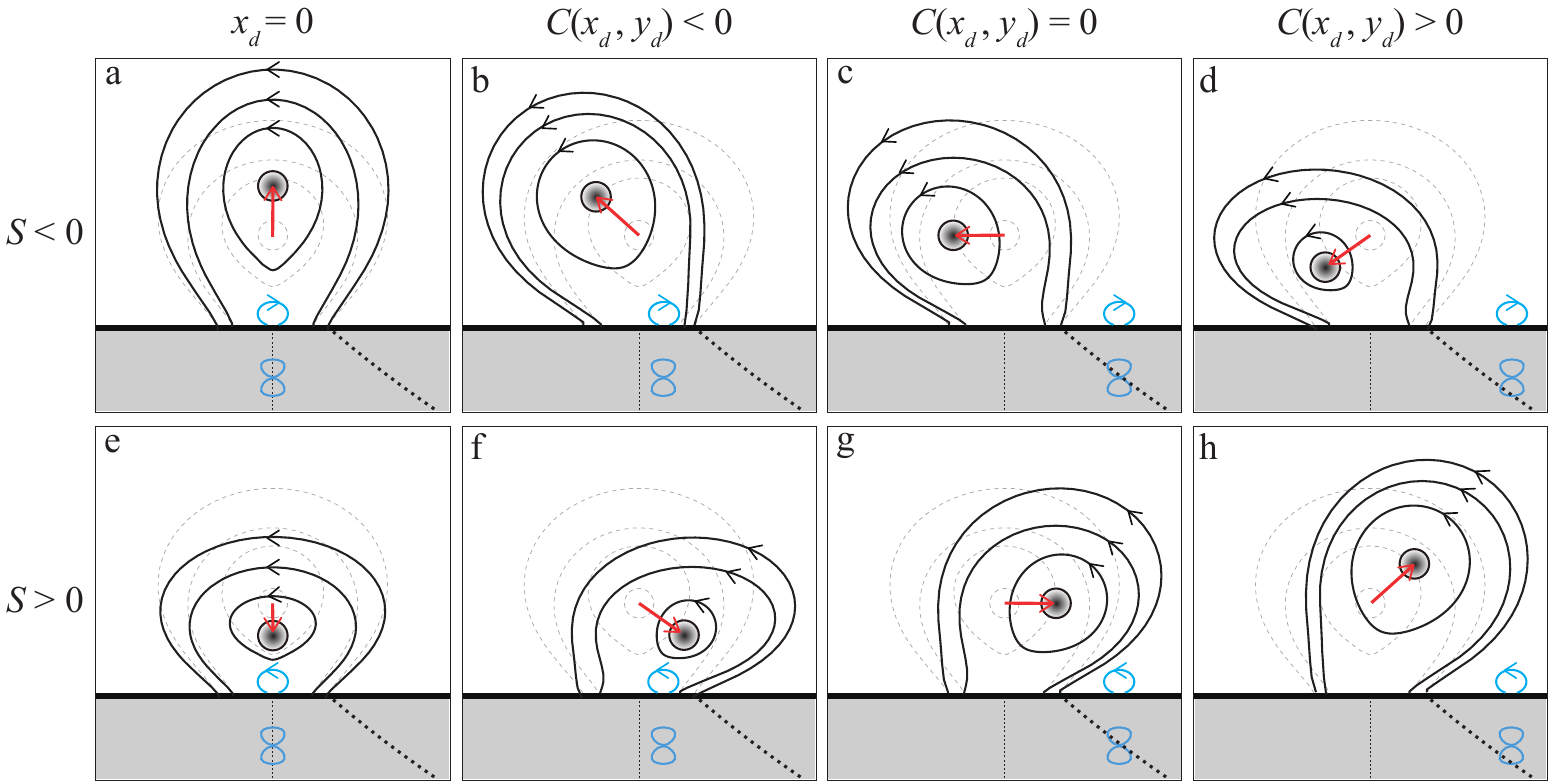}
\end{center}
\caption{Evolutionary tendencies of the FR as the new flux starts to emerge. The upper and the lower panels are for the NEF with $S<0$ and $S>0$, respectively. From left to right, four columns display magnetic configurations scenarios when Dipole 2 is positioned at different locations: $x_d=0$, $C(x_d,y_d)<0$, $C(x_d,y_d)=0$, and $C(x_d,y_d)>0$. Here, $C(x_d, y_d)$ is the channel function defined in equation (\ref{channel}). The gray dashed curves represent the initial magnetic configuration when the NEF is absent (i.e., at $S=0$), while the solid curves represent the configuration after being perturbed by the NEF. The cyan curves are for the magnetic field generated by the NEF. Two black dotted lines below the photosphere are determined by $x_d=0$ and $C(x_d,y_d)=0$. The red arrows indicate the direction of motion of the FR after the disturbance, and the shading inside the FR represents distribution of $J$.
\label{fig:trend}}
\end{figure}

\cite{2022ApJ...933..148C} performed a set of numerical experiments to look into the evolution in a similar system as a result of the NEF. They found that the early evolution may play a role of precursor in indicating the possible evolutionary behaviors of the disruptive magnetic configuration. To investigate how the system in equilibrium evolves as the new flux just commences to emerge, we start with $S=0$ (see magnetic structures marked as gray dashed lines in Figure \ref{fig:trend}) and the FR being at an equilibrium location with $x_h=0$, $y_h=1$, $J=1$, and $r_0=r_{00}$. Here, $r_{00}$ is the radius of the FR when $J=1$ and is utilized as a free parameter in this work, specifically set to $r_{00}=0.01$. While $r_{00}$ serves as a free parameter, it is intrinsically related to the internal equilibrium of the FR and, furthermore, affects the global equilibrium of the entire system (see also discussions in \citealt{1993ApJ...417..368I} and \citealt{1998ApJ...504.1006L}).

Any change in $\left|S\right|$ will introduce perturbation to the configuration. As \cite{2001JGR...10625053L} pointed out, on the other hand, the system does not necessarily lose the equilibrium immediately after such perturbation is invoked. Instead the configuration itself could find a new equilibrium state by adjusting the balance between the magnetic compression and the magnetic tension. Here, the magnetic compression tends to push the FR away from the Sun, and the magnetic tension tends to prevent the FR from moving away. Not until the critical point is reached will the configuration perturbed fails to maintain its equilibrium, and only then will the dynamic evolution of the system follow.

Taking derivatives of $x_h$ and $y_h$ on the basis of equations (\ref{externalx_eq}) through (\ref{forzenflux_eq}) with respective to $S$ at $S=0$ gives equations below (more details can be found in Appendix \ref{sec:appB}):
\begin{eqnarray}
\frac{\partial x_h}{\partial S}\bigg|_{S=0}&=&\frac{8x_d\left(1+y_d\right)}{\left[x_d^2+\left(1+y_d\right)^2\right]^2}, \label{partial x_full}\\
\frac{\partial y_h}{\partial S}\bigg|_{S=0}&=&
4\frac{\left[-y_d+1+2\ln\left(2/r_{00}\right)\right]x_d^2-\left[y_d+3+2\ln\left(2/r_{00}\right)\right]\left(y_d+1\right)^2}{\left[x_d^2+\left(1+y_d\right)^2\right]^2}, \label{partial y_full}
\end{eqnarray}
which govern the rate of changes in the FR positions due to the NEF at the very beginning. We noticed that when $x_d=0$, the FR exhibits no tendency to move horizontally (see Figures \ref{fig:trend}a and \ref{fig:trend}e), and that horizontal motion occurs as $x_{d}\neq 0$ (see the other panels in Figure \ref{fig:trend}).

Equation (\ref{partial x_full}) indicates that the displacement of the FR in $x$-direction always has the same sign of $S$, no matter where NEF appears (see also Figure \ref{fig:trend}). But that in $y$-direction shows a more complex behavior as indicated by equation (\ref{partial y_full}), depending on the sign of numerator at the right side of equation. Simple algebra indicates that the sign of the numerator is consistent with that of a channel function, $C(x_{d}, y_{d})$ that is defined as: 
\begin{eqnarray}
C(x_d, y_d)&=&x_d-\sqrt{\frac{y_d+3+2\ln\left(2/r_{00}\right)}{-y_d+1+2\ln\left(2/r_{00}\right)}}\left(y_d+1\right). \label{channel}
\end{eqnarray}
The channel function is valid only if the NEF is buried at a shallower region where $y_d < 1 + 2\ln(2/r_{00})$. As $y_d$ approaches $1 + 2\ln(2/r_{00})$, the boundary of the filament channel,  namely the $x_d$ determined by $C(x_d, y_d) = 0$, moves to infinity. Thus, the line $y_d = 1 + 2\ln(2/r_{00})$ in the $x_dy_d$-plane is an asymptote of the curve $C(x_d, y_d) = 0$. In the case of $r_{00}=0.01$ through out this work, this critical depth is $y_d^* = 11.6$. For those NEFs located deeper than $y_d = 1 + 2\ln(2/r_{00})$, the value of $C(x_d, y_d)$ is always negative.

Inspired by \cite{1995JGR...100.3355F} and \cite{2000ApJ...545..524C}, we have introduced the function $C(x_{d}, y_{d})$. \cite{1995JGR...100.3355F} and \cite{2000ApJ...545..524C} explored the FR evolution driven by NEFs located inside or outside the filament channel, and showed that the destiny of the evolution in the system more or less depends on the distance between the NEF and the pre-existing magnetic structures. Although \cite{2001JGR...10625053L} pointed out that dependence of the evolution in the system of interest on the NEF displays a more complex pattern than expected, no general description was given about how this dependence could be. 

After carefully looking into details of observations of \cite{1995JGR...100.3355F}, numerical experiments of \cite{2000ApJ...545..524C}, as well as analytic scenario of \cite{2001JGR...10625053L}, we realized the existence of a general description that outlines the boundary separating two distinct regions where the NEF leads to totally different behaviors and destinies of the FR evolution. It eventually turns out that this boundary could be determined by setting $C(x_{d}, y_{d}) = 0$.

As $S$ is located inside the filament channel and close to $M$, namely $C(x_{d},y_{d})<0$ (Figures \ref{fig:trend}b and \ref{fig:trend}f), NEF with the polarity opposite to that of $M$ tends to cause the FR to lose the equilibrium, and that with the same polarity tends to confine the FR at a more stable state. This scenario supports the conclusions of \cite{1995JGR...100.3355F} and \cite{2000ApJ...545..524C} regarding the dependence of the loss of equilibrium in the system on the relative orientation of $S$ and $M$. But as we will see later, the FR would (not) find new equilibrium position as $S> (<)$ 0 after losing the first equilibrium.

As $S$ is outside the filament channel and far from $M$, namely $C(x_{d},y_{d})>0$ (Figures \ref{fig:trend}d and \ref{fig:trend}h), the reverse situation occurs, namely the same polarity $S$ tends to cause the eruption, and the opposite polarity one does not, which is also consistent with the results of \cite{1995JGR...100.3355F} and \cite{2000ApJ...545..524C}. Consulting equation (\ref{partial y_full}), we realize that the above evolutionary behavior of the FR is governed not only by the polarity of the two sources, but by the location where the new source occurs as well. \cite{2001JGR...10625053L} qualitatively described the complexity of the evolutionary behavior of the FR, we are here giving a quantitative explanation.

Between the far and the close cases exists a special case in which the NEF is exactly located at the boundary of the filament channel (Figures \ref{fig:trend}c and \ref{fig:trend}g). When $S$ appears at this location, namely $C(x_{d}, y_{d})=0$, the FR tends to move horizontally during the evolution. Because this case is special, this might  explain why the CME moving horizontally is not popular, but is not impossible (see results of observations of \citealt{2010ApJ...722..329S} and of numerical experiments of \citealt{2022ApJ...933..148C}).

Before ending this part of work, we note here that varying $y_{d}$ and fixing $S$ is an alternative way of describing the NEF as indicated by equations (\ref{Bex}) and (\ref{Bey}). Duplications of the above algebra suggests that the evolutionary fashion of the system in response to the change in $y_{d}$ is simple and straightforward. The quasi-static evolution in the system eventually turns to dynamic in the case of $S>0$ following the catastrophe; and no catastrophe occurs, nor does the consequent dynamic evolution take place, as $S<0$.

\section{Evolutionary Behavior of the System in Equilibrium} \label{sec:2}

As mentioned earlier, solving equations (\ref{externalx_eq}) through (\ref{forzenflux_eq}) together determines an equilibrium position $(x_{h}, y_{h})$ of the FR for given $S$, $x_{d}$ and $y_{d}$. All these equilibrium positions, together with the corresponding combination of $S$, $x_{d}$ and $y_{d}$, constitute two equilibrium surfaces in 4-dimension spanned over spaces $(x_{h}, S, x_{d}, y_{d})$ and $(y_{h}, S, x_{d}, y_{d})$, respectively. The corresponding equilibrium surfaces for $J$ and $r_{0}$ are not independent, and could be deduced from the surfaces for $x_{h}$ and $y_{h}$.

\subsection{Equilibrium Surfaces of Multi-Fold and Critical Curves} \label{sec:2.1}

\begin{figure}[ht]
\begin{center}
\includegraphics[width=1.0\textwidth]{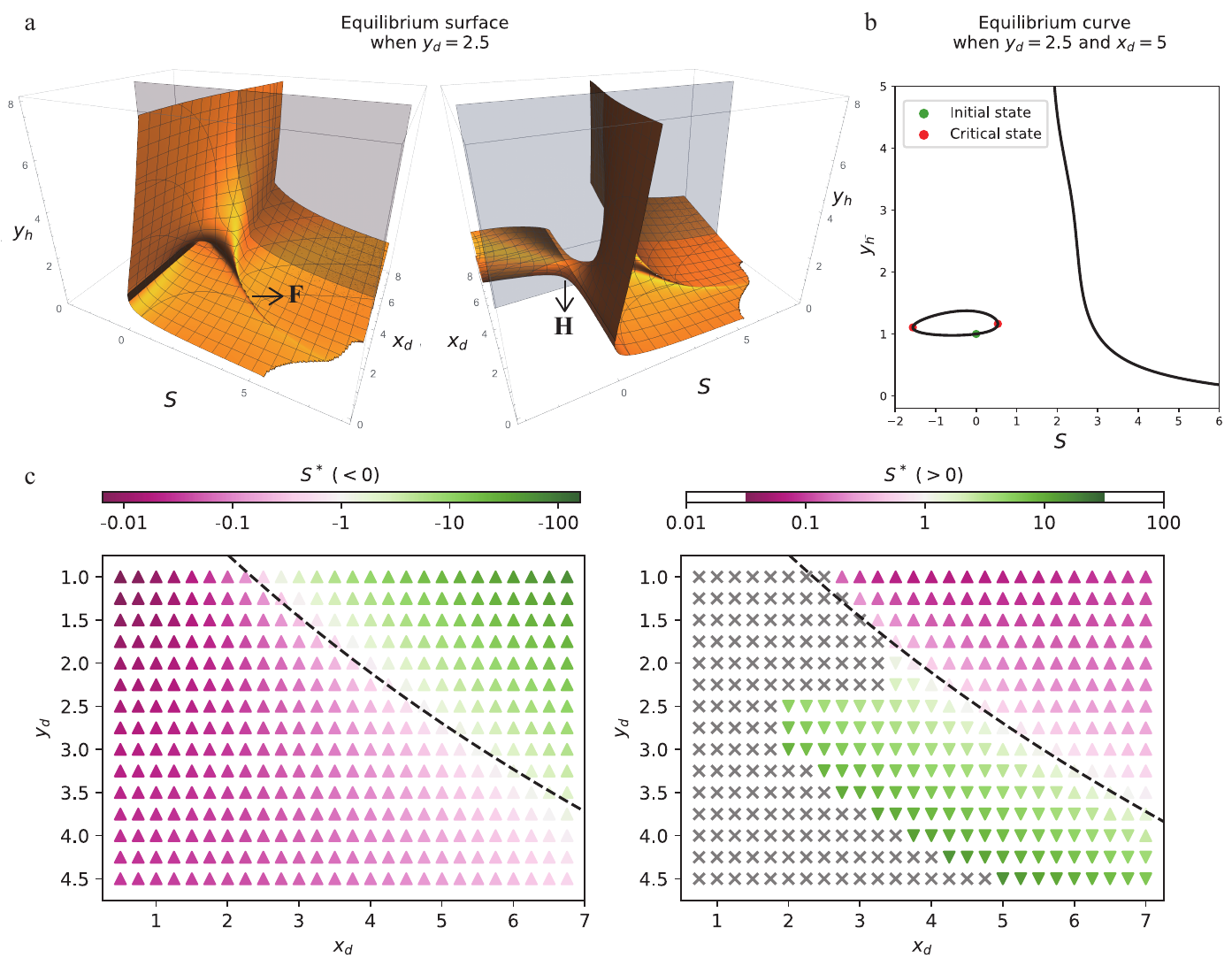}
\end{center}
\caption{(a) Equilibrium surface from two views shows variations in $y_h$, the equilibrium heights, as functions of both $S$ and $x_d$ with $y_d=2.5$, possessing two topological structures, i.e., the hole (H) and the fold (F). (b) Equilibrium curves of $y_h$ versus $S$ with $y_d=2.5$ and $x_d=5$. The green dot marks the FR's initial position before new flux emerges, while red dots indicate the critical states where the catastrophe occurs. (c) Distributions of $S^{*}$ in the space $(x_{d}, y_{d})$ with colors for $S^{*}$ values, regions filled with triangles (\(\bigtriangleup\)) for the combinations of parameters that yield catastrophe followed by a plausible eruption, regions filled with triangles up-side-down (\(\bigtriangledown\)) for cases of the catastrophe followed by the failed eruption, and that with crosses $(\times)$ for the case without catastrophe. The dashed line corresponds to $C(x_d, y_d)=0$. 
\label{fig:S}}
\end{figure}

To look into the equilibrium feature of the system governed by the solutions of equation (\ref{externalx_eq}) through (\ref{forzenflux_eq}), we need to project the 4D surface onto 3D space by fixing a variable, say $y_{d}$. Figure \ref{fig:S}(a) illustrates an equilibrium surface seen from two point of view with $y_d=2.5$, which describes the variation of $y_h$ versus $x_d$ and $S$, which manifests at least three features on a surface that indicates the occurrence of four types of catastrophe, namely no catastrophe,  the fold catastrophe, the cusp catastrophe, and the umbilic catastrophe. This implies that the complexity of the catastrophic behaviors of the system is significantly enhanced by just adding only one more source to the original configuration. Here the ``original configuration'' refers to the scenario where the NEF ($S$) is absent, leaving only a single dipole ($M$) that creates a symmetric background field. In this case, as discussed by \cite{2001JGR...10625053L} and earlier by \cite{1991ApJ...373..294F}, the system exhibits simple ``fold'' (``cusp'') catastrophe if a current sheet is (not) allowed to develop in the magnetic configuration prior to the loss of equilibrium (see also detailed discussions of \citealt{2001PhDT........13L} and \citealt{Lin2008theory}).

Further fixing $x_{d} = 5$ (see the gray plane in Figure \ref{fig:S}a) gives the equilibrium curve(s) of $y_{h}$ (and $x_{h}$ correspondingly) versus $S$ in a 2D space (see Figure \ref{fig:S}b). We see two equilibrium curves, a simple continuous curve without critical point (right one), and an oval with two critical points located at $(S=-1.56, y_{h}=1.11)$ and $(S=0.54, y_{h}=1.16)$, respectively (left one).

This indicates that, for the $S$ located at a medium depth $y_{d} = 2.5$, but is distant from $M$, say $x_{d}=5.0$, the catastrophe may take place no matter whether $S>0$ or $S<0$, namely either relative orientation of two sources could cause catastrophe. We also notice that the catastrophe may always occur in this case even if only small amount of magnetic flux emerges.

Complex structures of the equilibrium surface shown in Figure \ref{fig:S}a indicate various possibilities of the catastrophe that may happen as the system evolves in response to changes in $S$. But only the case that the evolution of the system in equilibrium starts with $S=0$ matches the scenario of the NEF studied in this work. So, for a NEF with the fixed location $(x_{d}, -y_{d})$, the gradual evolution in the system along the curve like that shown in Figure \ref{fig:S}b including both $S=0$ and the critical point $S=S^{*}$ would eventually end up with the catastrophe. Therefore, looking for $S^{*}$ and analyzing its dependence on the other parameters, like $x_{d}$ and $y_{d}$, are crucial for understanding detailed properties of a magnetic configuration evolving quasi-statically in response to the NEF, and for determining the critical state at which the quasi-static evolution turns into dynamic one.

In the previous section, we inferred the potential occurrence of a catastrophe based on the motion trend of the FR as the magnetic flux starts to emerge. Here, however, we directly investigate how the FR could lose equilibrium as a result of the NEF. Following the idea described above, we are able to find $S^{*}$ as a function of $x_{d}$ and $y_{d}$ by solving equation (\ref{Delta}). More than one solution could be expected, and some of them are of the NEF. To ensure that $S^{*}$ is on the curve or the surface that includes the point of $S=0$, a specific constraint needs imposing on solutions of equation (\ref{Delta}).

Start with a specific equilibrium curve that possesses at least one $S^{*}$ and includes $S=0$ for given $(x_{d}, y_{d})$, and slightly change $(x_{d}, y_{d})$, then we use root-finding algorithm to obtain another equilibrium curve that is close to the previous one, and includes both $S=0$ and an $S^{*}$. This approach ensures the smooth transition from one curve to another one. We repeatedly use this method to explore the whole domain of interest and acquire all the $S^{*}$ of the equilibrium curves that include the point of $S=0$. This guarantees that the critical state in the system could be reached by evolving the system in equilibrium gradually as a result of the NEF.

Figure \ref{fig:S}c displays the distribution of $S^{*}$ obtained by the method described above in space of $(x_{d}, y_{d})$. Left panel in Figure \ref{fig:S}c is for $S^{*}<0$, and right one for $S^{*}>0$. Triangles in both panels mark the region where $S^{*}$ exists, namely the catastrophe might occur, and crosses spread over the region where $S^{*}$ does not exist, namely the NEF occurring in this region does not lead to catastrophe.

In addition, the ordinary triangle means that the catastrophe may develop to the plausible eruption, and the up-side-down triangle implies that the motion of the FR after the catastrophe will be eventually stopped because a new equilibrium at another location exists for the FR.

Furthermore, the color bars in both panels represent the critical strength of $S$. We notice that the catastrophe is easy to occur as $S<0$ and $S$ is close to $M$ in the horizontal direction (namely, $x_{d}$ is small), and that the catastrophe becomes difficulty to happen as $x_{d}$ is large and both magnetic sources are far away from each other (see left panel in Figure \ref{fig:S}c). Qualitatively, the regions of the catastrophe-easy and the catastrophe-difficult roughly separate from each other by a narrow area around $S^{*}=1$. Interestingly, and yet intuitively, this corresponds to the line $C(x_d, y_d)=0$ in the $x_{d}y_{d}$-plane. As $S>0$, on the other hand, the opposite occurs with an area existing where no catastrophe happens (see right panel in Figure \ref{fig:S}c).

Consulting the results of \cite{1995JGR...100.3355F} and \cite{2000ApJ...545..524C}, we realize that the reconnection-favorable NEFs they studied share the same combination of location and polarity with the catastrophe-easy cases described here. Both are situated in regions of small $x_{d}$ and $S<0$ or large $x_{d}$ and $S>0$. In these scenarios, NEFs produce eruptive FRs, and only a weak NEF ($|S|<1$) is required for triggering the eruption.

As pointed out by \cite{2001JGR...10625053L}, the conclusions of \cite{1995JGR...100.3355F} and \cite{2000ApJ...545..524C} are true for several specific cases, and do not universally correct for all the cases related to the NEF. Specifically, our analysis highlights that the key distinction occurs in their reconnection-unfavorable case and our catastrophe-difficult case. These cases share the same combination of location and polarity, namely large $x_{d}$ and $S<0$ or small $x_{d}$ and $S>0$. The results of \cite{1995JGR...100.3355F} and \cite{2000ApJ...545..524C} indicated that the system is more likely to remain stable when the NEF is reconnection-unfavorable; and our analysis for catastrophe-difficult cases also support this conclusion when the NEF is weak ($|S|<1$). 

However, as the NEF gets stronger than the existing field ($|S|>1$), Figure \ref{fig:S}c demonstrates that the initially stable FR may either maintain its stability, or loses its equilibrium but eventually stabilize at a new position, or develops to a plausible eruption. This indicates a very complex evolutionary behavior of the coronal magnetic configuration driven by the NEF, which was noticed by \cite{2008ChA&A..32...56X} and \cite{2008SoPh..250...75Z}.

The present work demonstrates apparent diverse outcomes in catastrophe-difficult cases via introducing the channel function, which explicitly explained what had been missed by \cite{1995JGR...100.3355F} and \cite{2000ApJ...545..524C}. Overall, both \cite{2001JGR...10625053L} and the present work depicted a general scenario, in which a coronal magnetic configuration evolves in response to the NEF, and well cover the cases studied by \cite{1995JGR...100.3355F} and \cite{2000ApJ...545..524C}.

\subsection{Various Types of Evolution and the Corresponding Catastrophes} \label{sec:2.2}

\begin{figure}[ht]
\begin{center}
\includegraphics[width=0.85\textwidth]{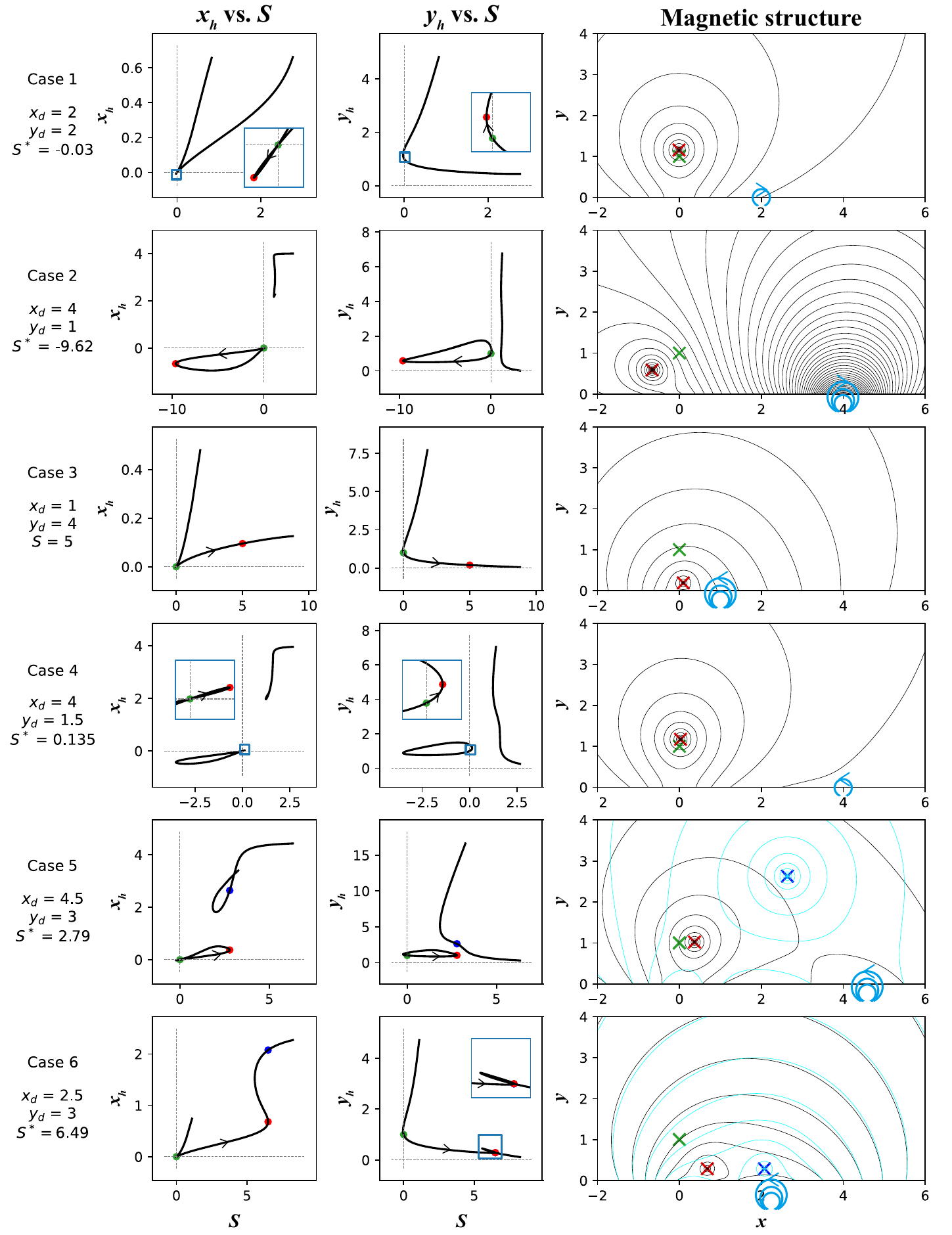}
\end{center}
\caption{Six cases of the FR evolution. Columns, from left to right, illustrate the evolution in $x_h$ with respect to $S$, the evolution in $y_h$, and the magnetic structure at the critical state, respectively. Each row illustrates a distinct case. Green dots on the curve and crosses in the magnetic structure represent initial positions of the FR, while red ones mark critical positions. In cases 5 and 6 (fifth and sixth rows), new equilibrium locations exist for the FR with the same $S$ at the critical point, so the FR might jump to the new location as indicated by the blue dots and crosses. The new magnetic structures after the catastrophe are plotted as cyan curves in right panels for cases 5 and 6. 
\label{fig:Five_Evo}}
\end{figure}

According to the distribution of $S^{*}$ displayed in Figure \ref{fig:S}c, we are able to extract six types of evolution in the system and the corresponding catastrophes. Figure \ref{fig:Five_Evo} displays the equilibrium curves for the six cases and the associated magnetic configurations. Although these plots are similar to those given by \cite{2001JGR...10625053L}, introduction of the channel function, $C$, gives us an approach to classifying systematically various essential features of the evolution in the coronal magnetic field driven by the NEF. This allows us to get the panoramic view of these features, and to figure out or forecast the destine of the evolution at the very beginning or in early stage of the evolution. Below we discuss these cases respectively on the basis of the plots shown in Figure \ref{fig:Five_Evo}.

Case 1: In this case, $C(x_d,y_d)<0$, the NEF appears in the region covered by the filament channel (see also Figure \ref{fig:trend}b or \ref{fig:trend}f). As the orientation of the NEF is opposite to that of the existing source, magnetic reconnection does not take place between the new magnetic field and that around the FR, but occurs between the new field and the background field passing around the top of the FR. This causes the confinement on the FR to be erased easily, and the catastrophe to take place even though only a small amount of new flux $(S^{*} = -0.03)$ emerges. Furthermore, Figure \ref{fig:Five_Evo} (case 1) shows that the shape of the equilibrium curves for both $x_{h}$ and $y_{h}$ is simple, and the critical point is of the fold catastrophe \citep{thom1972stabilite,poston1978catastrophe}. Thus the loss of equilibrium as a result of the catastrophe should develop to a plausible eruption since no new equilibrium exists for $S=-0.03$. Green and red crosses in the right panel of the figure mark the centers of the FR at the initial and the critical locations, respectively. The small distance between them indeed suggests that the catastrophe is easy to be triggered in this case, and the consequent motion of the FR just deviate from the $y$-direction slightly.

Case 2: This is the case of the umbilic catastrophe according to \cite{thom1972stabilite} and \cite{poston1978catastrophe}. In this case, the new flux emerges at a distant location from the pre-existing source, $C(x_d,y_d)>0$, which weakens its impact on the equilibrium of the FR. The existence of two sources causes the equilibrium curve to include two components, one corresponds to the NEF, and another one to that of the pre-existing source. The catastrophe is not easy to occur as a result of the NEF because the new source appears far from the FR as shown by panels in the second row of Figure \ref{fig:Five_Evo}. So a large amount of new flux $(|S^{*}| > 9)$ and significant deviation from the initial position of the FR is expected before the loss of equilibrium occurs, leading to a highly asymmetric critical magnetic structure. Consequently, we qualitatively infer that the FR will manifest apparent non-radial motion after the catastrophe. \cite{2021A&A...653L...2Z} reported an event that produced a CME that propagate almost horizontally in the early stage. Zhou et al. (2024) reported a similar event that was observed by ASO-S \citep{2019RAA....19..156G} recently. This event produced a CME that moved almost perpendicular to the radial direction at the very beginning.

Case 3 is simple and straightforward. Because the NEF has the same orientation $(S>0)$ as that of the pre-existing field, and $C(x_d,y_d)<0$ (see black crosses in Figure \ref{fig:S}c), the magnetic tension force on the FR is strengthened, and the confinement preventing the FR from losing equilibrium is consequently enhanced. The evolution in the system caused by the NEF in this case never leads to losing equilibrium. Instead the equilibrium becomes more stable.

Case 4: $x_{d}$ is the same as that in Case 2, and $C(x_d,y_d)>0$ as well. But two equilibrium curves exist in this case. However, only the left one with $S=0$ is for the evolutionary process driven by the NEF. These two curves do not share same values of $S$, so the loss of equilibrium occurring as a result of a certain type of evolution along one curve will not end at another curve for different type of evolution. The panel at right in the fourth row of Figure \ref{fig:Five_Evo} indicates that magnetic reconnection is easy to take place between the old and the new magnetic fields, the confinement on the FR is apparently weakened, and the loss of equilibrium of the FR then quickly occurs as $S^{*} = 0.135$. Since $S^{*}$ is small compared to $M$, deflection of the FR motion from the radial direction is not obvious, and a resulted CME propagating roughly in the radial direction could be expected.

We also notice that, on the other hand, the NEF with $S<0$ may also drive the system to evolve to the critical point. In this case, a larger amount of new flux is needed, and $S^{*}=-3.5$. Consequently, the location of the FR at the critical point is apparently deviates from the $y$-axis.  Namely, the value of $x_{h}$ at the critical point in this case is large, and the motion of the FR after the catastrophe starts from a location at which the force thrusting the FR outward possesses both $x$- and $y$-components. Therefore, the motion of the FR in non-radial direction could be expected.

Case 5: The NEF in this case appears at a location of both large $x_{d}$ and $y_{d}$ with $C(x_d,y_d)<0$ and results in a more complex umbilic catastrophe scenario than previous cases. This creates a strong background magnetic field in the region between two sources as $S>0$, and the equilibrium in the FR cannot be affected apparently as $S>0$ until a large amount of flux emerges. In this case, the FR is attracted to the new flux $S$, a relatively complicated magnetic configuration around the FR is created, and the confinement on the FR of the background is enhanced. 

We also notice that two equilibrium curves appear in the same domain such that two equilibrium locations of the FR exist for a given $S$ (see the left and the middle panels in the fifth row of Figure \ref{fig:Five_Evo}). One location could be reached via evolution in response to the NEF (see the lower closed curve), and another one belongs to different fashion of evolution (see the upper curve). The relative location of the two equilibrium curves indicates that the catastrophe triggered by the NEF may not be able to develop to a plausible eruption since the FR could find a new equilibrium at a higher and farther position. Namely, the loss of equilibrium occurring as a result of a certain type of evolution along one curve ends at another curve for different type of evolution. 

The reason why this can happen is because the consequent evolution following the catastrophe is stopped by the large-scale global background field that is not apparently disturbed by the catastrophe, and the confinement to the FR motion almost remains unchanged. The FR in this case behaves like an oscillator stuck to one end of a spring that is fixed to the wall with another end. The FR cannot escape from the spring if the spring is not broken although the FR might be moving quickly. Such an evolutionary feature of the FR has been confirmed by the numerical experiments of \cite{2022ApJ...933..148C}, and examples of failed eruptions were also reported by \cite{2003ApJ...595L.135J} in observations, by \cite{2000JGR...105.2375L} and \cite{2002ChJAA...2..539L} in analytic model, and by \cite{2005ApJ...630L..97T} in numerical experiments.

If the evolution in the system is driven by different mechanisms, on the other hand, the system will eventually lose the equilibrium at the critical point on the upper curve, and the catastrophe may develop to the eruption. Combining with the catastrophic scenario due to the NEF, this gives us an interesting evolutionary picture of the system. The NEF with $S>0$ drives the magnetic structure to deviate from the initially symmetric configuration with $S=0$, to evolve to the critical configuration with $S=S^{*}=2.79$, and the loss of equilibrium in the system thrusts the FR to the new equilibrium location specified by the upper curves in the left and the middle panels shown in the fifth row of Figure \ref{fig:Five_Evo}. Then, if $S$ decreases somehow, the evolution of the system in equilibrium will undergo along the upper curve, and eventually approach to another critical point with $S=S^{*}=1.83$. The consequent loss of equilibrium this time may develop to an eruption smoothly. This case could be tested numerically in the future as a follow-up of \cite{2022ApJ...933..148C}. We also expect this could be confirmed by observations as well.

If the NEF occurs in another way, say $S<0$, we find that the catastrophe may happen relatively easily. In this case, the FR is pushed to the region of $x<0$, and deviates from its initial equilibrium location in both $x$- and $y$-directions. We notice that the magnetic field has a simple configuration in the region of $x<0$, as the new flux pushes the FR away from the equilibrium in the $-x$-direction, the restoring force of the system may not be strong enough to pull the FR back to the initial position. Therefore, the NEF in this case could easily drive the FR to the critical location with $S^{*} = -0.23$, and then triggers the catastrophe. Since the catastrophe is triggered easily, and the consequent motion of the FR deviates from the radial direction slightly.

Case 6: This is another type of the failed eruption, in which the NEF is located at a position of small $x_d$ and $C(x_d,y_d)<0$. As $S>0$, a fold catastrophe is triggered at $S^*=6.49$, associated with a pair of critical points. Similar to Case 5, the catastrophe occurring in this case pushes the FR horizontally to the new equilibrium position (see the left and middle panels in the sixth row of Figure \ref{fig:Five_Evo}), demonstrating a motion pattern known as the transverse eruption \citep{2015SoPh..290.1703M}. 

Catastrophes occurring in Cases 5 and 6 with $S>0$ display similar consequences such that the FR is pushed rightward, and then reaches to the new equilibrium positions. The catastrophe in neither case would directly develop to a plausible eruption. But the difference between two cases exists as well. In Case 5, the new equilibrium position is on the curve that is for an alternative fashion of the evolution in the system. In Case 6, on the other hand, the new position is on the same equilibrium curve, showing similar feature to that studied by \cite{1991ApJ...373..294F} and \cite{1993ApJ...417..368I}. In both cases, however, the second catastrophe will occur again if $S$ decreases somehow after the FR reaches to the new position, and the consequent evolution would be able to develop to a plausible eruption that gives rise to a CME moving non-radially.

According to the rough classifications shown in Figure \ref{fig:S}c, we investigated the evolutionary behaviors of the magnetic configurations in response to the new flux that emerges in various fashions. Comparing with the observational results of \cite{1995JGR...100.3355F} and the numerical results of \cite{2000ApJ...545..524C}, we realized that magnetic reconnection is one of several issues that governs the catastrophic manifestations of the system. The location, orientation, and strength of the NEF are also able to determine the destination of the evolution. Among these issues, the strength of the NEF plays a role in bridging the reconnection-dominated and the non-reconnection-dominated catastrophes. For a weak NEF ($|S|<1$), reconnection is needed to destroy the original configuration to trigger the loss of equilibrium; for a strong NEF ($|S|>1$), on the other hand, the extra force imposed on the FR is enough to deform the initial magnetic configuration, and to trigger the loss of equilibrium. Additionally, a catastrophe triggered by a NEF with weak strength is expected to result in a CME propagating nearly in the radial direction, whereas a strong flux may produce a noticeably non-radial event. This could be tested and confirmed numerically in the future.

Generally speaking, Figure \ref{fig:S}c gives an overall scenario of the system evolution due to the NEF, but fails to reveal some details in depth. For example, in Case 5, the consequent evolution following the catastrophe due to the NEF may not continue smoothly because the FR would find the new equilibrium after losing the old equilibrium. When the system is in the new equilibrium state, if the NEF starts to weaken at a certain point, the system will evolve along the new equilibrium curve (see the left and the middle panels in the fifth row of Figure \ref{fig:Five_Evo}), and eventually loses the equilibrium as the new critical point is reached. Such a scenario might be rare, but not impossible. This indicates that different evolutionary fashions marked in the right panel in Figure \ref{fig:S}c could transfer among one another, and the locations of $S^{*}>0$ in Figure \ref{fig:Five_Evo} for Case 5  indicate that such a transfer is more likely to occur as $x_{d}$ and $y_{d}$ are located in the region near $C(x_d,y_d)=0$.

\section{Discussions} \label{sec:3}

\begin{figure}[ht]
\begin{center}
\includegraphics[width=0.75\textwidth]{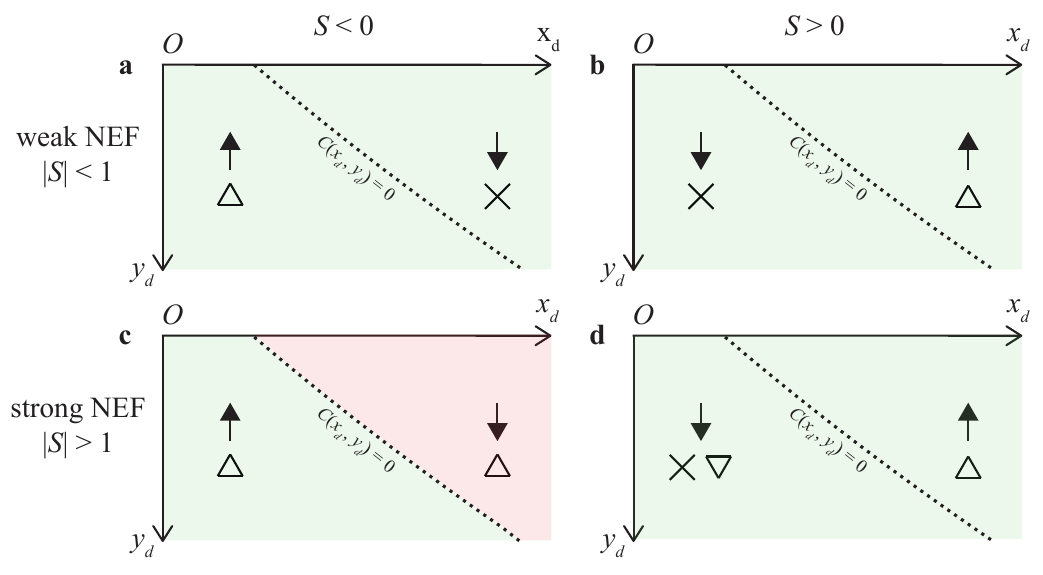}
\end{center}
\caption{The FR motion in the early stage and the consequent catastrophes driven by the NEF of varying polarities, strengths, and locations. Upward arrows denote the early ascent of the FR, while downward arrows indicate early descent. The meanings of triangles and crosses are the same as those in Figure \ref{fig:S}. The green region signifies that early FR ascent or descent is able to lead to either the presence or absence of CME, respectively, while the red region for the cases otherwise. Dotted lines are determined by $C(x_d,y_d)=0$.
\label{fig:early&final}}
\end{figure}

As a follow-up of \cite{2001JGR...10625053L}, we further analyze the evolution of a coronal magnetic configuration that includes an electric current-carrying FR in response to the NEF. Summarizing the results of \cite{2001JGR...10625053L} and new discoveries obtained in the present work, we find one criterion regarding the initial behaviors and the consequent manifestations of the magnetic structure governed by the NEF: $C(x_d,y_d)=0$. Figure \ref{fig:early&final} outlines two regions in the parameter space, $(x_{d}, y_{d})$, separated by the dotted line for $C(x_d,y_d)=0$. The arrows specify the motion direction of the FR in the $y$-direction during the early stage of evolution as the NEF just turns on.

On the basis of Figure \ref{fig:S}c, Figure \ref{fig:early&final} displays in a more apparent fashion how the destination of the system evolution depends on combinations of parameters. In the case of a weak NEF, say $|S|<1$ (Figures \ref{fig:early&final}a and \ref{fig:early&final}b), the early motion pattern of the FR could determine the destination of the system evolution for any $(x_{d}, y_{d})$. We notice that the upward (downward) motion of the FR initially driven by the NEF located in any region, no matter what the polarity of the NEF is, will (not) end up with the catastrophe eventually. This result is consistent with the discoveries of \cite{2001ApJ...559..452Z,2012NatCo...3..747Z} and \cite{2023ApJ...954L..47C} such that the ``slow-rise'' of filaments could serve as a precursor of CMEs. 

When the emerging field is strong, i.e., $\left|S\right| >1$ (Figures \ref{fig:early&final}c and \ref{fig:early&final}d), the situation becomes fairly complex. In this scenario, the correlation of the final evolutionary behavior of the system to the initial behavior is roughly the same as in the case of $|S|<1$ with two exceptions. One is for NEFs with $S<0$ located in the region of $C(x_d,y_d)>0$, the initially descending FR leads to loss of equilibrium as $|S|$ increases; and another one for NEFs with $S>0$ and $C(x_d,y_d)<0$, even a strong NEF could not create successful eruptions.

The above results indicate that if $C(x_d,y_d)<0$ the fashion of the system evolution in the late stage is consistent with that in the early stage; but if the NEF appears far from the FR, with $C(x_d,y_d)>0$, the consistence may or may not exist, depending on the polarity and strength of the NEF. In addition, the early rising of an FR always leads to an eruption given sufficient emerging flux, whereas the early descent results in either an eruption or a non-eruption. This suggests that the evolutionary behavior of a magnetic configuration including the NEF is more complex than expected although the overall magnetic structure is simple (e.g., see also discussions of \citealt{2001JGR...10625053L}).

Although the magnetic configurations investigated in this study could be considered simple,  valuable insight of the fundamental physics governing the interaction between the NEF and the pre-existing coronal magnetic structure has been obtained. Construction of the channel function, $C$, allows us to outline various regions in the parameter domain that accounts for the magnetic structure of interest, and to predict the associated evolutionary behaviors. The transition between the two magnetic configurations of Case 5 in Figure \ref{fig:Five_Evo} presents an example of transition between two configurations in equilibrium by the catastrophe, and numerical experiments by \cite{2022ApJ...933..148C} demonstrated that such a transition is possible.

\begin{figure}[ht]
\begin{center}
\includegraphics[width=0.6\textwidth]{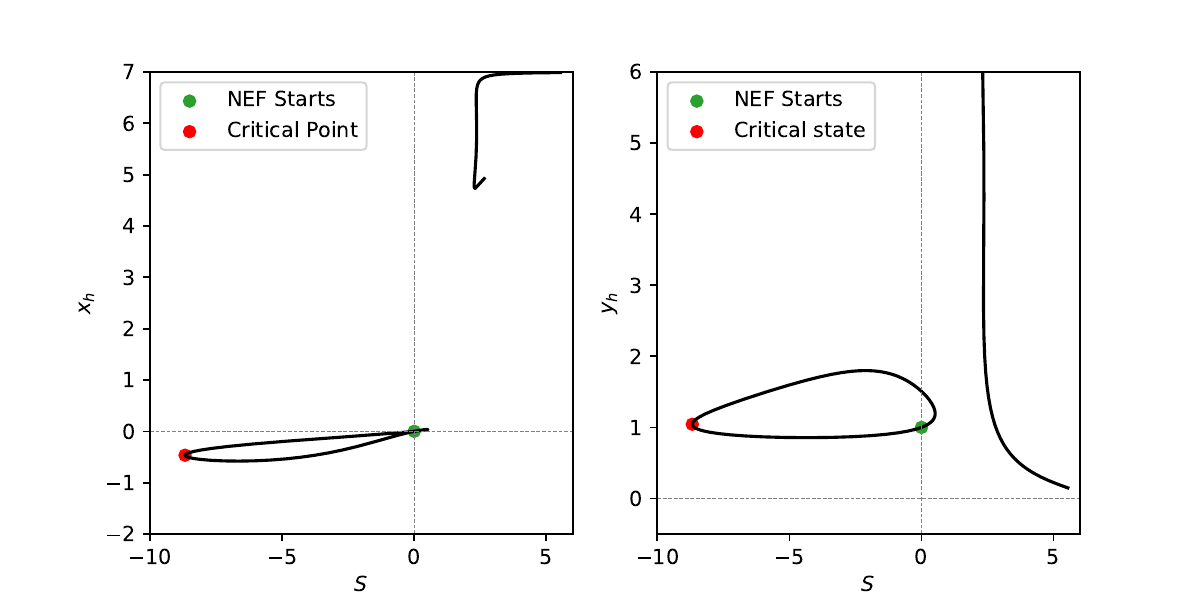}
\end{center}
\caption{Counterpart of the Case 2 in Figure \ref{fig:Five_Evo} for $r_{00} = 0.05$, $x_{d} = 7$, and $y_{d} = 2.5$. The NEF occurs as $S$ changes in the negative direction.
\label{fig:curve}}
\end{figure}

To give one more example that the dynamic evolution in the magnetic configuration after the catastrophe predicted by the analytic solution could be duplicated numerically, we re-do the calculations of \cite{2022ApJ...933..148C} for the Case 2 shown in Figure \ref{fig:Five_Evo} with $r_{00} = 0.05$, $x_{d} = 7$, and $y_{d} = 2.5$, and the other setups remaining unchanged. The reason to set $r_{00} = 0.05$, instead of $r_{00} = 0.01$, is to suppress the numerical diffusion in calculations, and that to take values of $x_{d}$ and $y_{d}$ different from those for the Case 2 in Figure 4 is to leave enough space between two equilibrium curves in order to avoid crosstalk between different evolutionary fashions after the catastrophe takes place.

\begin{figure}[ht]
\begin{center}
\includegraphics[width=0.6\textwidth]{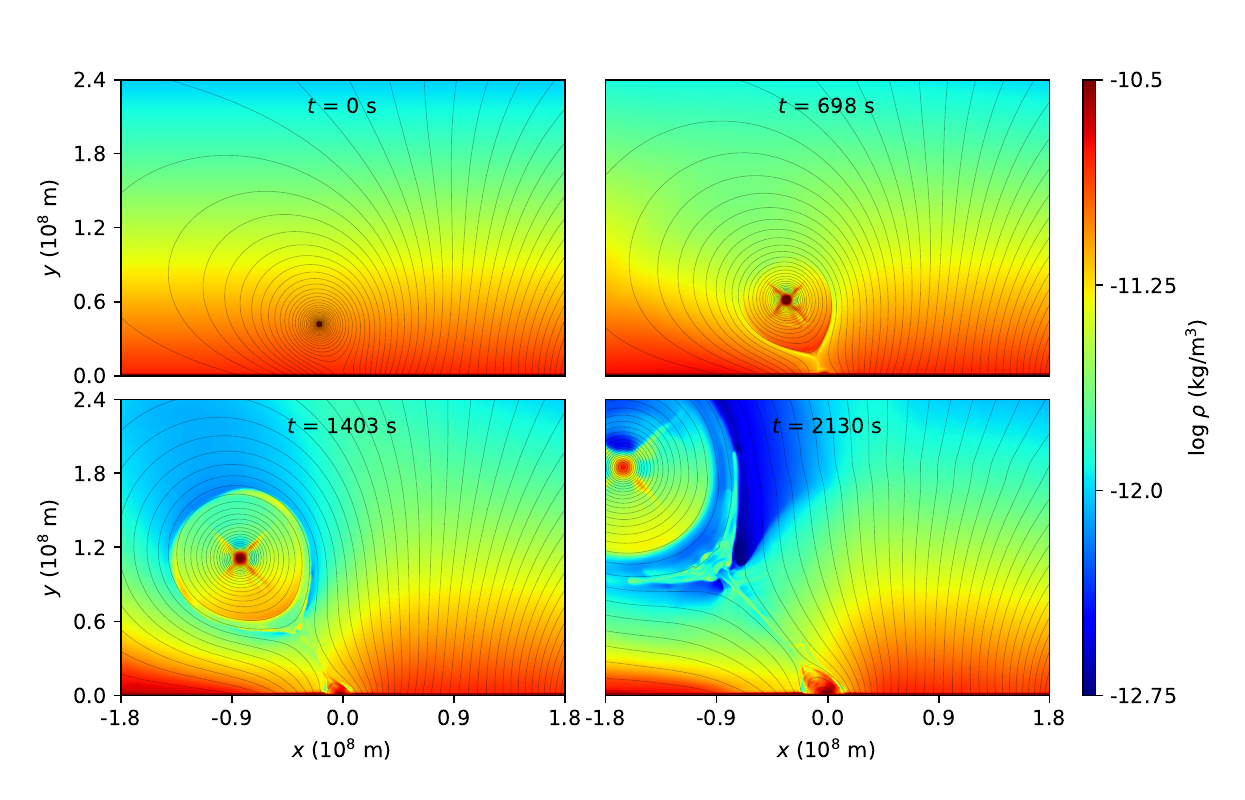}
\end{center}
\caption{Evolutions of the magnetic field (continuous curves) and the associated plasma distributions (colorful shadows) after the catastrophe described by Figure \ref{fig:curve}. The initial magnetic configuration ($t = 0$~s) corresponds to the critical point in Figure \ref{fig:curve} (red dots).
\label{fig:evo}}
\end{figure}

Figure \ref{fig:curve} is the counterpart of the Case 2 in Figure \ref{fig:Five_Evo} for $r_{00} = 0.05$, $x_{d} = 7$, and $y_{d} = 2.5$. It shows the same feature as that of the latter although the values of $r_{00}$, $x_{d}$, and $y_{d}$ in the two cases are different. The equilibrium curves shown in Figure 6 suggest the loss of equilibrium in the relevant magnetic configuration would eventually develop to a plausible eruption. Figure \ref{fig:evo} displays a set of time sequences of numerical results for the same magnetic structure (continuous curves) and the associated distributions of the plasma density around (colorful shadows) after the loss of equilibrium. The fine structures of the current sheet behind the FR (Figure \ref{fig:evo} at $t=2130$~s) indicate that magnetic reconnection occurring in the current sheet is turbulent. This, combining with that shown by \cite{2022ApJ...933..148C}, indicates that the results and the associated conclusions obtained from the analytic solution could be indeed duplicated or validated by numerical experiments. More detailed investigations of the same topic will be conducted numerically in the future.

\section{Conclusions} \label{sec:Conclusion}

We looked into detailed evolutionary behaviors of a coronal magnetic structure that includes an electric current-carrying FR as a result of the new emerging flux. Following the practice of \cite{2001JGR...10625053L}, the background magnetic field results from two magnetic dipoles embedded inside the photosphere. The location, $(0, -d)$, and the strength, $M$, of one dipole is fixed for the basic background field, and another one of location $(x_{d}, -y_{d})$ and strength $S$ is used to model the NEF. The emerging process is described by changing $S$ from $S=0$ for given $(x_{d}, -y_{d})$. The initial magnetic configuration with $S=0$ is symmetric about the $y$-axis. As the new flux starts to emerge, the symmetry in the system is inevitably destroyed.

Evolutionary behaviors of the system are governed by strength, orientation, and location of $S$. Changes in the equilibrium location of the FR, $(x_{h}, y_{h})$, as functions of these parameters constitute two hyper-surfaces $(S, x_{d}, y_{d}, x_{h})$ and $(S, x_{d}, y_{d}, y_{h})$, respectively, in the 4-dimensional parameter space. Fix any of $S$, $x_{d}$, and $y_{d}$, two 3-dimensional surfaces could be obtained for $x_{h}$ and $y_{h}$, respectively, and equilibrium curves in 2-dimension are further given as $y_{d}$ is fixed. Six types of equilibrium curves may be deduced: one describes the case of no catastrophe, and the others correspond to five types of catastrophe at the end of the quasi-static evolution in the system: fold with successful eruption, fold with failed eruption, cusp, the umbilic with two equilibrium curves without overlap, and the umbilic with two equilibrium curves between which overlap exists.

Comparing with \cite{2001JGR...10625053L}, we analyzed the impact of the NEF on evolution in the coronal magnetic configuration in a more systematic fashion by introducing the channel function, $C(x_{d}, y_{d})$. This function provides a way to look into how the NEF’s location, strength, and orientation governs the evolutionary behavior of the system, and allows us to determine whether the evolution in the system would end up with a catastrophe in response to the NEF.

By looking into magnetic configurations with the NEF being invoked and the consequent evolution of the system along the equilibrium curves mentioned above, we notice a line $C(x_d,y_d)=0$ existing on the $x_{d}y_{d}$-plane, which determines several motion patterns of the FR in the early stage (e.g., see Figure \ref{fig:trend}). For $S<0$ $(S>0)$, these patterns include purely upward (downward) as $x_{d}=0$, toward up and left (down and right) as $C(x_d,y_d)<0$, toward purely left (right) as $C(x_d,y_d)=0$, and toward down and left (up and right) as $C(x_d,y_d)>0$, respectively. Because of the symmetry in the system about the $y$-axis, we just need to study the cases of $x_{d} > 0$.

We note here that the existence of the channel function $C$ provides us a way to analyze systematically the overall behavior of the coronal magnetic configuration driven by the NEF. With the help of $C$, we are capable of distinguishing regions in the parameter space over which specific types of catastrophe occur, and even figuring out the destine of a given type of evolution from very beginning or in the early stage.

We also found that the evolution of the magnetic configuration in the late stage may not always be consistent with that in the early stage. This is because the strength, in addition to the location and orientation, of the NEF is also an important issue that governs the evolution of the system. In the case of a negative NEF $(S<0)$ near the FR, or a positive NEF $(S>0)$ far from the FR, a small amount of the new flux is required for driving the system to the critical state and invoking the catastrophe, namely $|S^{*}|<1$. Here, $S^*$ is the critical amount of flux required for a catastrophe. In this case, the relative orientation of the new flux and the pre-existing magnetic field determines whether the catastrophe occurs such that the catastrophe occurs if the relative orientation favors for reconnection, and no catastrophe otherwise. This result is consistent with the observations of \cite{1995JGR...100.3355F} and the numerical results of \cite{2000ApJ...545..524C}.

In the case of a positive NEF $(S>0)$ close to the FR, or a negative NEF $(S<0)$ far away from the FR, a large amount of new flux is needed to drive the system to the critical state $(|S^{*}|>1)$, or the catastrophe never occurs, $S^{*}$ does not exist. This suggests that $|S^{*}|=1$ is a special value of the critical strength that separates different fashions of evolution in the system and the corresponding catastrophe. Further studies indicate that the loss of equilibrium in the system caused by the NEF will occurs with $|S^{*}|=1$ if the location of the NEF $(x_{d}, y_{d})$ satisfies $C(x_d,y_d)=0$.

The solution curve to $C(x_d,y_d)=0$ divides the $x_{d}y_{d}$-plane into two areas, and for the given location of the NEF, we are able to determine the destination of the system evolution according to the manifestation of the system in the early stage. For an NEF appear near the FR, namely $C(x_d,y_d)<0$, upward motion of the FR in the early stage will eventually evolve into an eruption, and the early downward motion of the FR results in a more stable configuration in equilibrium or a failed eruption. For an NEF appears at large distance from the FR $C(x_d,y_d)>0$, the situation becomes complex, and the evolutionary behavior of the system is governed by both the polarity and the location of the NEF. 




\vspace{\baselineskip} 
\noindent This work was supported by the National Key R\&D Program of China No. 2022YFF0503800, NSFC grants 12073073 and 11933009, grants associated with the Yunnan Revitalization Talent Support Program, the Foundation of the Chinese Academy of Sciences (Light of West China Program), the Yunling Scholar Project of Yunnan Province and the Yunnan Province Scientist Workshop of Solar Physics, and grants 202101AT070018 and 2019FB005 associated with the Applied Basic Research of Yunnan Province. M.Z. acknowledges the support by the NSFC grants 12273107 and U2031141. Support was also from the Yunnan Key Laboratory of Solar Physics and Space Science (202205AG070009).
We benefit from the discussions of the ISSI-BJ Team ``Solar eruptions: preparing for the next generation multi-waveband coronagraphs''. Calculations in this work were carried out on the cluster in the Computational Solar Physics Laboratory of Yunnan Observatories.



\appendix

\section{Equilibrium Equations}\label{sec:appA}

According to \cite{2001JGR...10625053L}, the external magnetic field (the total magnetic field subtracted by the magnetic field generated by the FR itself) at the center of the FR is:
\begin{eqnarray}
B_{ex}(x_h,y_h)&=&\frac{2I_0}{cd} \left\{\frac{2M\left[x_h^{2}-\left(y_h+1\right)^{2}\right]}{\left[x_h^{2}+\left(y_h+1\right)^{2}\right]^{2}}
  +\frac{2S\left[\left(x_h-x_d\right)^{2}-\left(y_h+y_d\right)^{2}\right]}{\left[\left(x_h-x_d\right)^{2}+\left(y_h+y_d\right)^{2}\right]^{2}}+\frac{J}{2y_h}\right\}\label{Bex},\\
B_{ey}(x_h,y_h)&=&\frac{2I_0}{cd} \left\{\frac{4M x_h\left(y_h+1\right)}{\left[x_h^{2}+\left(y_h+1\right)^{2}\right]^{2}}
  +\frac{4S\left(x_h-x_d\right)\left(y_h+y_d\right)}{\left[\left(x_h-x_d\right)^{2}+\left(y_h+y_d\right)^{2}\right]^{2}}\right\}\label{Bey},
\end{eqnarray}
and the Lorentz force acting on the FR is given by:
\begin{eqnarray}
{\bf{F}}=\frac{I_0J}{c}{\bf{z}}\times{\bf{B_e}}\label{F},
\end{eqnarray}
where $\bf{z}$ is the unit vector in the $z$-direction, and the two components of $\bf{B_e}$ are given by equations (\ref{Bex}) and (\ref{Bey}).

Four equations governing the equilibrium state of the FR are (e.g., see also \citealt{2001JGR...10625053L}):
\begin{eqnarray}
\frac{2M\left[x_h^{2}-\left(y_h+1\right)^{2}\right]}{\left[x_h^{2}+\left(y_h+1\right)^{2}\right]^{2}}
  +\frac{2S\left[\left(x_h-x_d\right)^{2}-\left(y_h+y_d\right)^{2}\right]}{\left[\left(x_h-x_d\right)^{2}+\left(y_h+y_d\right)^{2}\right]^{2}}+\frac{J}{2y_h}&=&0,\label{externalx_eq}\\
\frac{Mx_h\left(y_h+1\right)}{\left[x_h^{2}+\left(y_h+1\right)^{2}\right]^{2}}
  +\frac{S\left(x_h-x_d\right)\left(y_h+y_d\right)}{\left[\left(x_h-x_d\right)^{2}+\left(y_h+y_d\right)^{2}\right]^{2}}&=&0,\label{externaly_eq}\\
r_0&=&\frac{r_{00}}{J},\label{internal_eq}\\
J\ln\left(\frac{2y_hJ}{r_{00}}\right)+\frac{2M\left(y_h+1\right)}{\left[x_h^{2}+\left(y_h+1\right)^{2}\right]}+
\frac{2S\left(y_h+y_d\right)}{\left(x_h-x_d\right)^{2}+\left(y_h+y_d\right)^{2}}&=&\ln\left(\frac{2}{r_{00}}\right)+1. \label{forzenflux_eq}
\end{eqnarray}
Global equilibrium in the system requires $\bold{F}$ in equation (\ref{F}) to vanish, which yields that $B_{ex}$ and $B_{ey}$ in equations (\ref{Bex}) and (\ref{Bey}) equal to zero, resulting in equations (\ref{externalx_eq}) and (\ref{externaly_eq}), respectively. The internal equilibrium of the FR leads to equation (\ref{internal_eq}) that relates the radius of the FR, $r_{0}$, to the current density inside the FR, $J$, with $r_{00}$ being the value of $r_{0}$ as $J=1$. Finally, equation (\ref{forzenflux_eq}) is deduced from the frozen-flux condition on the FR surface.

The free energy stored in the system, $E_{h}$, is equal to the work done to move the FR from the equilibrium location, $(x_{h}, y_{h})$, to infinity:
\begin{equation}
-\mathrm{d}E_{h} = \mathbf{F} \cdot \mathrm{d}\mathbf{r}, \label{dE}
\end{equation}
where $\bf{F}$ is given by equation (\ref{F}), and ${\bf{r}} = (x, y)$ is the position vector of the FR center. Substituting equation (\ref{F}) into (\ref{dE}), and completing the integration leads to:
\begin{equation}
E_h = \left( \frac{I_0}{c} \right)^2 J^2 \left[ \ln \left( \frac{2 y_h J}{r_{00}} \right) + \frac{1}{2} \right]. \label{Eh}
\end{equation}

\section{Approach to Introducing the Channel Function, $C$}\label{sec:appB}

The deduction of the channel function $C(x_d,y_d)$ is given in this part of work. We start with evaluating $\partial x_h/ \partial S$ and $\partial y_h/ \partial S$ at $S=0$ (see Equations (\ref{partial x_full}) and (\ref{partial y_full})). Four functions of $x_{h}$, $y_{h}$, $r_{0}$, $J$, and $S$ for given $x_{d}$ and $y_{d}$ are defined according to Equations (\ref{externalx_eq}) through (\ref{forzenflux_eq}):
\begin{eqnarray}
f_1&=&\frac{2M\left[x_h^{2}-\left(y_h+1\right)^{2}\right]}{\left[x_h^{2}+\left(y_h+1\right)^{2}\right]^{2}}
  +\frac{2S\left[\left(x_h-x_d\right)^{2}-\left(y_h+y_d\right)^{2}\right]}{\left[\left(x_h-x_d\right)^{2}+\left(y_h+y_d\right)^{2}\right]^{2}}+\frac{J}{2y_h},\label{f1}\\
f_2&=&\frac{Mx_h\left(y_h+1\right)}{\left[x_h^{2}+\left(y_h+1\right)^{2}\right]^{2}}
  +\frac{S\left(x_h-x_d\right)\left(y_h+y_d\right)}{\left[\left(x_h-x_d\right)^{2}+\left(y_h+y_d\right)^{2}\right]^{2}},\label{f2}\\
f_3&=&r_0-\frac{r_{00}}{J},\label{f3}\\
f_4&=&J\ln\left(\frac{2y_hJ}{r_{00}}\right)+\frac{2M\left(y_h+1\right)}{\left[x_h^{2}+\left(y_h+1\right)^{2}\right]}+
\frac{2S\left(y_h+y_d\right)}{\left(x_h-x_d\right)^{2}+\left(y_h+y_d\right)^{2}}-\ln\left(\frac{2}{r_{00}}\right)-1. \label{f4}
\end{eqnarray}

By taking the derivative of $f_i$ with respect to $S$ and applying the chain rule, we obtain：
\begin{eqnarray}
    \frac{df_i}{dS}=\frac{\partial f_i}{\partial x_h} \frac{\partial x_h}{\partial S} + \frac{\partial f_i}{\partial y_h} \frac{\partial y_h}{\partial S} + \frac{\partial f_i}{\partial r_0} \frac{\partial r_0}{\partial S} + \frac{\partial f_i}{\partial J} \frac{\partial J}{\partial S} + \frac{\partial f_i}{\partial S} = 0, \quad i = 1, 2, 3, 4.\label{dfds}
\end{eqnarray}
The matrix form of Equation (\ref{dfds}) is:
\begin{equation}
A
\begin{bmatrix}
    \frac{\partial x_h}{\partial S} \\
    \frac{\partial y_h}{\partial S} \\
    \frac{\partial r_0}{\partial S} \\
    \frac{\partial J}{\partial S}
\end{bmatrix}
+
 \begin{bmatrix}
    \frac{\partial f_1}{\partial S} \\
    \frac{\partial f_2}{\partial S} \\
    \frac{\partial f_3}{\partial S} \\
    \frac{\partial f_4}{\partial S}
\end{bmatrix}
=0,\label{dfds_mat}
\end{equation}
where $A$ is the Jacobian matrix:
\begin{equation}
A = \begin{bmatrix}
    \frac{\partial f_1}{\partial x_h} & \frac{\partial f_1}{\partial y_h} & \frac{\partial f_1}{\partial r_0} & \frac{\partial f_1}{\partial J} \\
    \frac{\partial f_2}{\partial x_h} & \frac{\partial f_2}{\partial y_h} & \frac{\partial f_2}{\partial r_0} & \frac{\partial f_2}{\partial J} \\
    \frac{\partial f_3}{\partial x_h} & \frac{\partial f_3}{\partial y_h} & \frac{\partial f_3}{\partial r_0} & \frac{\partial f_3}{\partial J} \\
    \frac{\partial f_4}{\partial x_h} & \frac{\partial f_4}{\partial y_h} & \frac{\partial f_4}{\partial r_0} & \frac{\partial f_4}{\partial J}
\end{bmatrix}. \label{Jacobi}
\end{equation}
Thus, the partial derivatives related to the motion of the FR are:
\begin{equation}
\begin{bmatrix}
    \frac{\partial x_h}{\partial S} \\
    \frac{\partial y_h}{\partial S} \\
    \frac{\partial r_0}{\partial S} \\
    \frac{\partial J}{\partial S}
\end{bmatrix}
= - A^{-1}
\begin{bmatrix}
    \frac{\partial f_1}{\partial S} \\
    \frac{\partial f_2}{\partial S} \\
    \frac{\partial f_3}{\partial S} \\
    \frac{\partial f_4}{\partial S}
\end{bmatrix}.\label{C}
\end{equation}

Substituting Equations (\ref{f1}) through (\ref{f4}) into Equation (\ref{Jacobi}) (noting that at $S=0$, the equilibrium state of the FR is characterized by $x_h=0$, $y_h=1$, $J=1$, $r_0=r_{00}$), we obtain $A^{-1}$:
\begin{equation}
A^{-1}\bigg|_{S=0} = \begin{bmatrix}
    0 & 8 & 0 & 0 \\
    -4 - 4\ln\left(\frac{2}{r_{00}}\right) & 0 & 0 & 2 \\
    -2r_{00} & 0 & 1 & 0 \\
    2 & 0 & 0 & 0
\end{bmatrix}.\label{inverseA}
\end{equation}
Then, we calculate the vector on the right-hand side of Equation (\ref{C}):
\begin{equation}
\begin{bmatrix} 
    \frac{\partial f_1}{\partial S} \\ 
    \frac{\partial f_2}{\partial S} \\ 
    \frac{\partial f_3}{\partial S} \\ 
    \frac{\partial f_4}{\partial S}
\end{bmatrix}\bigg|_{S=0}
=
\begin{bmatrix} 
    \frac{2\left[x_d^2-(1+y_d)^2 \right]}{\left[x_d^2+(1+y_d)^2 \right]^2} \\ 
    -\frac{x_d(1+y_d)}{\left[x_d^2+(1+y_d)^2 \right]^2} \\ 
    0 \\ 
    \frac{2(1+y_d)}{x_d^2+(1+y_d)^2} 
\end{bmatrix}.\label{B}
\end{equation}

By substituting Equations (\ref{inverseA}) and (\ref{B}) into Equation (\ref{C}), we finally obtain the partial derivatives expected:
\begin{eqnarray}
\frac{\partial x_h}{\partial S}\bigg|_{S=0} &=& \frac{8x_d(1+y_d)}{\left[x_d^2+(1+y_d)^2\right]^2}, \label{APPpartial x_full} \\
\frac{\partial y_h}{\partial S}\bigg|_{S=0} &=& -4\frac{\left[y_d-1-2\ln\left(\frac{2}{r_{00}}\right)\right]x_d^2+\left[y_d+3+2\ln\left(\frac{2}{r_{00}}\right)\right](y_d+1)^2}{\left[x_d^2+(1+y_d)^2\right]^2}, \label{APPpartial y_full} \\
\frac{\partial J}{\partial S}\bigg|_{S=0} &=& -\frac{4\left[x_d^2-(1+y_d)^2\right]}{\left[x_d^2+(1+y_d)^2\right]^2}, \label{APPpartial J_full} \\
\frac{\partial r_0}{\partial S}\bigg|_{S=0} &=& \frac{4r_{00}\left[x_d^2-(1+y_d)^2\right]}{\left[x_d^2+(1+y_d)^2\right]^2}. \label{APPpartial r_full}
\end{eqnarray}

We notice that equations (\ref{partial x_full}) and (\ref{partial y_full}) are hence obtained, and they are identified with equations (\ref{APPpartial x_full}) and (\ref{APPpartial y_full}), respectively. Finally, the channel function, $C(x_d, y_d)$, in equation (\ref{channel}) is thus defined according to equation (\ref{APPpartial y_full}) or (\ref{partial y_full}).


\bibliography{cyh}{}
\bibliographystyle{aasjournal}


\end{CJK*}
\end{document}